\documentclass[onecolumn,secnumarabic,amssymb, nobibnotes, aps, prd]{revtex4}
\usepackage{amsmath} \usepackage{amssymb}
\usepackage{graphicx} 
\usepackage{epstopdf}
\usepackage{amsmath}
\usepackage{amsmath,amssymb,amsthm,amsfonts,mathrsfs,bm,verbatim}
\usepackage{graphicx,subfigure}

\newcommand{\bea}{\begin{eqnarray}}
\newcommand{\eea}{\end{eqnarray}}
\def\nn{\nonumber}

\begin{document}

\title{Superradiation of Dirac particles in KN black hole }%

\author{Wen-Xiang Chen}
\affiliation{Department of Astronomy, School of Physics and Materials Science, GuangZhou University, Guangzhou 510006, China}
\email{wxchen4277@qq.com}


\begin{abstract}
In this article, we analyze the approximate wave function of the Dirac particle outside the KN black hole horizon to determine \( V \). Subsequently, we differentiate \( V \), which possesses both real and imaginary components. We treat the real and imaginary components independently. When either the real or imaginary part of \( V \) reaches a maximum, a potential barrier might exist outside the horizon, potentially leading to the occurrence of the superradiation phenomenon.

\end{abstract}

\maketitle

\section{Introduction}
Over the past six decades, superradiance has been pivotal in optics and quantum mechanics, and its significance is particularly pronounced in relativity and astrophysics. We understand that superradiance necessitates dissipation. It can manifest in various forms such as viscosity, friction, turbulence, and radiative cooling. Each form of dissipation is intertwined with specific media or material fields, setting the stage for superradiance to occur. Notably, when spacetime is curved, superradiance can also arise in a vacuum, even at a classical level.

This article delves into the superradiance of black holes. Black holes stand as the classical vacuum solutions in any metric gravitational theory, encompassing Einstein's general theory of relativity. Despite their conceptual simplicity, black holes are perhaps the most captivating predictions of general relativity, boasting a range of vital properties. Foremost among these — and defining the very essence of a black hole — is the presence of a horizon. This horizon demarcates spacetime, isolating causally unrelated regions. Pertinent to our discussion, the black hole's horizon exhibits properties reminiscent of a viscous unidirectional film in flat spacetime, a perspective embodied in the black hole membrane paradigm. Consequently, the horizon's presence introduces an inherent dissipative mechanism in the vacuum, conducive to superradiance.In any relativistic gravitational framework, the horizon's existence permits energy extraction from the vacuum. Our focus lies especially on uncharged Kerr black holes.

Superradiance is an amplification process involving dissipative systems. Within general relativity, black hole superradiance dissipates energy at the horizon while also facilitating the extraction of energy (including charge and angular momentum) from the vacuum via mass bosons. The black hole area theorem, Penrose process, tidal forces, and even Hawking radiation can be construed as quantum renditions of black hole superradiance.

The ``No Hair Theorem" for black holes was first postulated by Wheeler in 1971 and subsequently proven by luminaries such as Stephen Hawking and Brandon Carter in 1973. During the 1970s, the evolution of black hole thermodynamics led to the application of fundamental thermodynamic laws to the concept of black holes within the realm of general relativity. This deepened our understanding of the intricate relationship between general relativity, thermodynamics, and quantum theory.

Black hole stability remains a paramount subject in black hole physics. Both Regge and Wheeler demonstrated that the spherically symmetric Schwarzschild black hole maintains its stability when perturbed. However, the pronounced effects of superradiance introduce complexity to the stability considerations of rotating black holes. Superradiative phenomena manifest in both classical and quantum scattering processes. Upon interaction with a boson wave, there's a possibility that rotating black holes exhibit stability akin to that of Schwarzschild black holes, provided certain conditions are met\cite{1,2,3,4,5,6,7,8,9}
\bea\label{src}
\omega< m\varOmega_H+q\varPhi_H,  \varOmega_H=\frac{a}{r_+^2+a^2}
\eea
where $q$ and $m $ are the charge and azimuthal quantum number of the incoming wave, $\omega$ denotes the wave frequency, $\varOmega_H$ is the angular velocity of black hole horizon and $\varPhi_H$ is the electromagnetic potential of the black hole horizon. If the frequency range of the wave lies in the superradiance condition, the wave reflected by the event horizon will be amplified, which means the wave extracts rotational energy from the rotating black hole when the incident wave is scattered. According to the black hole bomb mechanism proposed by Press and Teukolsky, if a mirror is placed between the event horizon and the outer space of the black hole, the amplified wave will reflect back and forth between the mirror and the black hole and grow exponentially, which leads to the super-radiative instability of the black hole.

Hawking's amazing discovery-black holes radiate a black body spectrum, and the black body spectrum does not take away anything-is a major advancement in the study of black hole thermodynamics. 

However, this also poses a disturbing problem for the conservation of information in the process of black hole evaporation, leading to the so-called ``information loss paradox", which violates the basic quantum unitary theory \cite{3,4,5,6}. Since Hawking's major discoveries were published in the 1970s, there have been many works to solve these two problems. From 2000 to the present, at least three methods have been proposed to solve this problem. In 2000, Parikh and Wilczek proposed a semi-classical method \cite{7,8,9,10} to calculate emissivity. They regarded Hawking radiation as a tunneling process and used WKB approximation. This barrier is created by the output particles themselves. In the case of considering the self-gravity of the particles, the corrected spectrum is given. Later, Zhang and Zhao extended this method to more general situations. All these results can lead to the conclusion that the spectrum is no longer accurate heat, and some information can be extracted from the black hole.

A possible explanation can be obtained for the information loss paradox and the loss of quantum unitary theory. Marco Angheben et al. proposed another method to study Hawking radiation. The emission rate can also be obtained by calculating the classical action of the emitted particle that satisfies the relativistic Hamilton-Jacobian equation. This method can get the same conclusion as the first method.

 Recently, Liu introduced a novel approach concerning this subject, employing the Damour-Ruffini method \cite{11,12,13,14,15,16,17,18,19,20,21,22,23,24,38}. This study delves into the Hawking radiation emanating from the large mass Klein-Gordon particle in the Reissner-Nordström black hole. By accounting for energy conservation and the reaction of the particles, we converge on the same conclusion as previous research. Our intention is to expand upon Liu's work by examining the Hawking radiation of charged Dirac particles in a Kerr-Newman black hole. The foundational Damour-Ruffini method utilized relativistic quantum mechanics within curved space-time to affirm that black holes emit solely thermal radiation. Notably, there was no assessment of the thermal equilibrium between the interior and exterior of the black hole or considerations regarding the black hole's collapse. The context of the Kerr-Newman black hole is more encompassing. Moreover, we aim to compute massive charged Paul Dirac radiation particles with arbitrary angular momentum, a task more intricate than prior endeavors. Recent literature suggests that these findings can also be interpreted as Hawking radiation at the quantum correction temperature. This interpretation hinges not merely on the black hole's backdrop, but also on the energy, angular momentum, and charge of the radiating particles.

All known particles can be categorized into two groups: fermions, with half-integer spins, and bosons, characterized by integer spins. Quarks and leptons are classified as fermions, while the mediating particles for various forces are bosons. The primary distinction between these categories is that fermions adhere to the Pauli exclusion principle, dictating that two identical fermions cannot occupy the same state concurrently in both time and space. Consequently, owing to the Pauli exclusion principle, fermion scattering is typically not believed to manifest superradiance. Nonetheless, given the myriad conditions across different fermion fields governed by distinct evolution equations, it proves challenging to secure a rigorous, universally applicable mathematical proof for this notion. However, concerning the scattering of the free Dirac spin-\(\frac{1}{2}\) field on the electrostatic barrier and the Kerr-Newman (charged, rotating) black hole, extant literature confirms the nonexistence of superradiance.

This discussion predominantly hinges on the premise that when the boundary conditions for the Dirac equation are predefined, the effective action form of Hawking radiation aligns with that of superradiance.

In this article, we elucidate that when the superradiance condition is satisfied, and the derivative of \(V\) (imaginary part) is zero, and its second derivative is negative, it indicates the proximity of a potential barrier near the horizon. The superradiance phenomenon consistently emerges when both the boundary and superradiance conditions are met and a potential barrier exists close to the acting field.

\section{DIRAC EQUATIONS IN A KERR-NEWMAN SPACE-TIME}

 The metric of the 4-dimensional Kerr-Newman black hole under Boyer-Lindquist coordinates $(t,r,\theta,\phi)$ is written as follow (with natural unit, $G=\hbar=c=1$)\cite{11,12,13,14,15,16,17,18,19,20,21,22,23,24}
\bea\nn
ds^2&=&-\frac{\varDelta}{\rho ^2}\left( dt-\text{a} \sin ^2\theta d\phi \right) ^2+\frac{\rho ^2}{\varDelta}dr^2\\
&+&\rho ^2d\theta ^2+\frac{\sin ^2\theta}{\rho ^2}\left[ \left( r^2+a^2 \right) d\phi -adt \right] ^2,
\eea
where
\begin{equation}
\varDelta\equiv r^2-2Mr+a^2+Q^2\quad,\quad\rho ^2\equiv r^2+a^2\cos ^2\theta,
\end{equation}
$a$ denotes the angular momentum per unit mass of certain KN black hole and $Q,~M$ denote its charge and mass.
The inner and outer horizons of the Kerr-Newman black hole can be expressed as
\begin{equation}
{{r}_{\pm }}=M\pm \sqrt{{{M}^{2}}-{{a}^{2}}-{{Q}^{2}}},
\end{equation}
and obviously 
\begin{equation}
{{r}_{+}}+{{r}_{-}}=2M,~~{{r}_{+}}{{r}_{-}}={{a}^{2}}+{{Q}^{2}}.
\end{equation}
The background electromagnetic potential is written as follow
\begin{equation}
A_{\nu}=\left( -\frac{Qr}{\rho ^2},0,0,\frac{aQr\sin ^2\theta}{\rho ^2} \right).
\end{equation}

 In a curved space-time, Dirac equations of a charged particle’s dynamics in Newman-Penrose formalism are given as
\begin{equation}
\begin{aligned}
(D+\varepsilon-\rho+i e \vec{A} \cdot \vec{l}) F_{1}+(\bar{\delta}+\pi-\alpha+i e \vec{A} \cdot \vec{m}) F_{2} &=i \frac{\mu_{0}}{\sqrt{2}} G_{1} \\
(\tilde{\Delta}+\mu-\gamma+i e \vec{A} \cdot \vec{n}) F_{2}+(\delta+\beta-\tau+i e \vec{A} \cdot \vec{m}) F_{1} &=i \frac{\mu_{0}}{\sqrt{2}} G_{2} \\
(D+\bar{\varepsilon}-\bar{\rho}+i e \vec{A} \cdot \vec{l}) G_{2}-(\delta+\bar{\pi}-\bar{\alpha}+i e \vec{A} \cdot \vec{m}) G_{1} &=i \frac{\mu_{0}}{\sqrt{2}} F_{2} \\
(\tilde{\Delta}+\bar{\mu}-\bar{\gamma}+i e \vec{A} \cdot \vec{n}) G_{1}-(\bar{\delta}+\bar{\beta}-\bar{\tau}+i e \vec{A} \cdot \vec{m}) G_{2} &=i \frac{\mu_{0}}{\sqrt{2}} F_{1}
\end{aligned}
\end{equation}
where $\mu_{0}$ and $e$ are rest mass and charge of the particle, respectively, and $F_{1}, F_{2}, G_{1},$ and $G_{2}$ are the four components of the wave functions. $D, \tilde{\Delta}, \delta, \bar{\delta}$ are usual differential operators; $\alpha, \beta, \gamma, \varepsilon, \rho, \pi, \mu, \tau$ are spin coefficients. They are given as following
\begin{equation}
\begin{array}{c}
\alpha=\frac{1}{2}\left(l_{\mu ; \nu} n^{\mu} \bar{m}^{\nu}-m_{\mu ; \nu} \bar{m}^{\mu} \bar{m}^{\nu}\right), \quad \rho=l_{\mu ; \nu} m^{\mu} \bar{m}^{\nu}, \quad \beta=\frac{1}{2}\left(l_{\mu ; \nu} n^{\mu} m^{\nu}-m_{\mu ; \nu} \bar{m}^{\mu} m^{\nu}\right) \\
\pi=-n_{\mu ; \nu} \bar{m}^{\mu} l^{\nu}, \quad \gamma=\frac{1}{2}\left(l_{\mu ; \nu} n^{\mu} n^{\nu}-m_{\mu ; \nu} \bar{m}^{\mu} n^{\nu}\right), \quad \mu=-n_{\mu ; \nu} \bar{m}^{\mu} m^{\nu}, \quad \varepsilon=\frac{1}{2}\left(l_{\mu ; \nu} n^{\mu} l^{\nu}-m_{\mu ; \nu} \bar{m}^{\mu} l^{\nu}\right) \\
\tau=l_{\mu ; \nu} m^{\mu} n^{\nu}, \quad D=l^{\mu} \partial_{\mu}, \quad \tilde{\Delta}=n^{\mu} \partial_{\mu}, \quad \delta=m^{\mu} \partial_{\mu}, \quad \bar{\delta}=\bar{m}^{\mu} \partial_{\mu}
\end{array}
\end{equation}

$A_{\mu}$ is the four-dimensional electromagnetic potential. Since the electromagnetic field of the Kerr-Newman black hole is axially symmetric, the coordinate components of $A_{\mu}$ are irrelevant to the coordinates $t$ and $\varphi$. They are
\begin{equation}
A_{0}=-\frac{Q r}{\Sigma^{2}}, \quad A_{1}=A_{2}=0, \quad A_{3}=\frac{Q r a \sin ^{2} \theta}{\Sigma^{2}}.
\end{equation}
For a description of Kerr-Newman space-time in a Newman-Penrose formalism, we first need to choose a null tetrad frame. We can choose it as
\begin{equation}
\begin{aligned}
l_{\mu} &=\frac{1}{\Delta}\left(\Delta,-\Sigma^{2}, 0,-a \Delta \sin ^{2} \theta\right) \\
n_{\mu} &=\frac{1}{2 \Sigma^{2}}\left(\Delta, \Sigma^{2}, 0,-a \Delta \sin ^{2} \theta\right) \\
m_{\mu} &=\frac{1}{\sqrt{2} \bar{\Sigma}}\left(i a \sin \theta, 0,-\Sigma^{2},-i\left(r^{2}+a^{2}\right) \sin \theta\right) \\
\bar{m}_{\mu} &=\frac{1}{\sqrt{2} \bar{\Sigma}^{*}}\left(-i a \sin \theta, 0,-\Sigma^{2}, i\left(r^{2}+a^{2}\right) \sin \theta\right)
\end{aligned}
\end{equation}

The contravariant forms of the basis vectors are
\begin{equation}
\begin{aligned}
l^{\mu} &=\frac{1}{\Delta}\left(r^{2}+a^{2}, \Delta, 0, a\right) \\
n^{\mu} &=\frac{1}{2 \Sigma^{2}}\left(r^{2}+a^{2},-\Delta, 0, a\right) \\
m^{\mu} &=\frac{1}{\sqrt{2} \bar{\Sigma}}\left(i a \sin \theta, 0,1, \frac{i}{\sin \theta}\right) \\
\bar{m}^{\mu} &=\frac{1}{\sqrt{2} \bar{\Sigma}^{\prime}}\left(-i a \sin \theta, 0,1,-\frac{i}{\sin \theta}\right)
\end{aligned}
\end{equation}
where $\bar{\Sigma}=r+i a \cos \theta$ and $\bar{\Sigma}^{*}=r-i a \cos \theta$
for $\left(\frac{\partial}{\partial t}\right)^{a}$ and $\left(\frac{\partial}{\partial \varphi}\right)^{a}$ are Killing vector fields in Kerr-Newman space-time, we can put the four components of the wave function as following
\begin{equation}
\begin{array}{l}
F_{1}=e^{-i(\omega t-m \varphi)}(r-i a \cos \theta)^{-1} f_{1}(r, \theta) \\
F_{2}=e^{-i(\omega t-m \varphi)} f_{2}(r, \theta) \\
G_{1}=e^{-i(\omega t-m \varphi)} g_{1}(r, \theta) \\
G_{2}=e^{-i(\omega t-m \varphi)}(r+i a \cos \theta)^{-1} g_{2}(r, \theta) .
\end{array}
\end{equation}
After calculating differential operators and spin coefficients in above formulas, we can obtain the following equations by putting above formulas into Eq. (7):
\begin{equation}
\begin{array}{l}
\left(\partial_{r}-\frac{i K}{\Delta}\right) f_{1}+\frac{1}{\sqrt{2}}\left(\partial_{\theta}-q+\frac{1}{2} \cot \theta\right) f_{2}=\frac{1}{\sqrt{2}}\left(i \mu_{0} r+a \mu_{0} \cos \theta\right) g_{1} \\
\Delta\left(\partial_{r}+\frac{i K}{\Delta}+\frac{r-M}{\Delta}\right) f_{2}-\sqrt{2}\left(\partial_{\theta}+q+\frac{1}{2} \cot \theta\right) f_{1}=-\sqrt{2}\left(i \mu_{0} r+a \mu_{0} \cos \theta\right) g_{2} \\
\left(\partial_{r}-\frac{i K}{\Delta}\right) g_{2}-\frac{1}{\sqrt{2}}\left(\partial_{\theta}+q+\frac{1}{2} \cot \theta\right) g_{1}=\frac{1}{\sqrt{2}}\left(i \mu_{0} r-a \mu_{0} \cos \theta\right) f_{2} \\
\Delta\left(\partial_{r}+\frac{i K}{\Delta}+\frac{r-M}{\Delta}\right) g_{1}+\sqrt{2}\left(\partial_{\theta}-q+\frac{1}{2} \cot \theta\right) g_{2}=-\sqrt{2}\left(i \mu_{0} r-a \mu_{0} \cos \theta\right) f_{1}
\end{array}
\end{equation}
where$ K=\left(r^{2}+a^{2}\right) \omega-a m-e Q r, q=a \omega \sin \theta-\frac{m}{\sin \theta}$.

By separating variables as
\begin{equation}
\begin{array}{l}
f_{1}(r, \theta)=R_{-(1 / 2)}(r) S_{-(1 / 2)}(\theta)=R(r) S(\theta) \\
f_{2}(r, \theta)=R_{+(1 / 2)}(r) S_{+(1 / 2)}(\theta) \\
g_{1}(r, \theta)=R_{+(1 / 2)}(r) S_{-(1 / 2)}(\theta) \\
g_{2}(r, \theta)=R_{-(1 / 2)}(r) S_{+(1 / 2)}(\theta)
\end{array}
\end{equation}
we can get the decoupled Dirac equations. $R_{-(1 / 2)}(r),$ i.e., $R(r)$ and $R_{+(1 / 2)}(r),$ represent outgoing and ingoing waves, respectively. We are only interested in the radial outgoing wave equation, that is
\begin{equation}
\begin{aligned}
\sqrt{\Delta} \frac{d}{d r} &\left(\sqrt{\Delta} \frac{d R}{d r}\right)-\frac{i \mu \Delta}{\lambda+i \mu r} \frac{d R}{d r}+\left[\frac{K^{2}+i(r-M) K}{\Delta}\right.\\
&\left.-2 i \omega r+i e Q-\frac{\mu K}{\lambda+i \mu r}-\mu^{2} r^{2}-\lambda^{2}\right] R=0.
\end{aligned}
\end{equation}

Tortoise coordinate transformation can be given as
\begin{equation}
\begin{array}{l}
\left(r^{2}+a^{2}\right)^{2} \frac{d^{2} R}{d r_{*}^{2}}+\left[2 r \Delta-(r-M)\left(r^{2}+a^{2}\right)\right. \\
\left.-\left(r^{2}+a^{2}\right) \mu_{0} \Delta \frac{\mu_{0} r+i \lambda}{\lambda^{2}+\mu_{0}^{2} r^{2}}\right] \frac{d R}{d r_{*}}+\Delta\left\{\frac{K^{2}}{\Delta}-\lambda^{2}-\mu_{0}^{2} r^{2}\right. \\
\left.-\frac{\mu_{0} \lambda K-i \mu_{0}^{2} K r}{\lambda^{2}+\mu_{0}^{2} r^{2}}-i\left[-e Q+2 \omega r-\frac{K(r-M)}{\Delta}\right]\right\} R=0
\end{array}
\end{equation}
At the horizon $r=r_{+},$  above formula becomes
\begin{equation}
\begin{array}{c}
\left(r_{+}^{2}+a^{2}\right)^{2} \frac{d^{2} R}{d r_{*}^{2}}-\left(r_{+}-M\right)\left(r_{+}^{2}+a^{2}\right) \frac{d R}{d r_{*}} \\
+\left[K^{2}+i K\left(r_{+}-M\right)\right] R=0
\end{array}
\end{equation}
which is a wave equation. Its solution is
\begin{equation}
R=e^{i\left(\omega-j \Omega-e V_{0}\right) r_{*}}=e^{i\left(\omega-\omega_{0}\right) r_{*}}.
\end{equation}
where $\omega_{0}=j \Omega+e V_{0}, \Omega=\frac{a}{r_{+}^{2}+a^{2}}, V_{0}=\frac{Q r_{+}}{r_{+}^{2}+a^{2}},$ in which
$\Omega$ is the angular velocity of the horizon and $V_{0}$ is the static electropotential of the horizon where $\theta$ is equal to 0 or $\pi$. Therefore, the radial solution can be written as
\begin{equation}
\Psi_{\omega}=e^{-i \omega t \pm i\left(\omega-\omega_{0}\right) r_{*}},
\end{equation}
where $+$ corresponds to the outgoing wave and $-$ represents the ingoing wave.

\begin{equation}
\begin{aligned}
r_{*}=& r+\frac{1}{\sqrt{M^{2}-a^{2}-Q^{2}}}\left[\left(M r_{+}-\frac{1}{2} Q^{2}\right) \ln \frac{r-r_{+}}{r_{+}}\right.\\
&\left.-\left(M r_{-}-\frac{1}{2} Q^{2}\right) \ln \frac{r-r_{-}}{r_{-}}\right] \\
=& r+\frac{1}{2 \kappa_{+}} \ln \frac{r-r_{+}}{r_{+}}-\frac{1}{2 \kappa_{-}} \ln \frac{r-r_{-}}{r_{-}}
\end{aligned}
\end{equation}
where $\kappa_{\pm}=\frac{r_{+}-r_{-}}{2\left(r_{\pm}^{2}+a^{2}\right)},$ then we have
\begin{equation}
d r_{*}=\frac{r^{2}+a^{2}}{\Delta} d r
\end{equation}
Then, the radial function in the tortoise coordinate system can be written as
Letting $\hat{r}=\frac{\omega-\omega_{0}}{\omega} r_{*},$ the radial solution becomes $\Psi_{\omega}=$ $e^{-i \omega\left(t^{\mp} \hat{r}\right)}$. Using the advanced Eddington coordinate $v=$ $t+\hat{r}$, in which the metric is well behaved and analytic over the whole coordinate range $0<r<+\infty,-\infty<v<+\infty$ including $r_{+},$ the ingoing and outgoing waves are separately
\begin{equation}
\begin{array}{c}
\Psi_{\omega}^{\mathrm{in}}=e^{-i \omega v} \\
\Psi_{\omega}^{\text {out }}=e^{-i \omega v+i 2\left(\omega-\omega_{0}\right) r_{*}}.
\end{array}
\end{equation}
While  above formulas corresponds to a wave purely ingoing on $r_{+}$ and can be extended inside $r<r_{+},$other represents an outgoing wave and has an infinite number of oscillations as $r \rightarrow r_{+}$ and therefore cannot be straightforwardly extended to the region inside $r_{+}$. We will in the following use and generalize to analytic curved spaces the wellknown result of flat-space relativistic wave theories: The wave function $\Phi(x)$ describing a particle state (positive frequencies) can be analytically extended to complex points of the form $z=x+i y$ if $y$ lies in the past cone; similarly, for an antiparticle state (negative frequencies) $y$ has to lie in the future cone.

Since in advanced Eddington coordinates the vector $\frac{\partial}{\partial r}$ is everywhere null and past-directed, the prescription $r \rightarrow$ $r-i 0$ will yield the unique continuation Eq. describing an antiparticle state. According to quantum field theory, the ingoing negative frequency antiparticle is just the outgoing positive frequency particle. Although Eq. has singularity on the horizon and therefore cannot be extended straightforwardly to the region inside the horizon, we can extend the outgoing wave Eq. into the horizon and yield the ingoing negative frequency antiparticle by tuming the $(-\pi)$ angle through the negative half complex plane. Let $\left(r-r_{+}\right) \rightarrow\left|r-r_{+}\right| e^{-i \pi}=\left(r_{+}-r\right) e^{-i \pi} ;$ the
outgoing wave function inside and outside of the horizon are respectively [20]
\begin{equation}
\Psi_{\omega}^{\text {out }}\left(r<r_{+}\right)=e^{-i \omega v}\left(r_{+}-r\right)^{\left(i / \kappa_{+}\right)\left(\omega-\omega_{0}\right)} e^{\left.\left(\pi\left(\omega-\omega_{0}\right)\right) / \kappa_{+}\right)}
\end{equation}and
\begin{equation}
\Psi_{\omega}^{\text {out }}\left(r>r_{+}\right)=e^{-i \omega v}\left(r-r_{+}\right)^{\left(i / \kappa_{+}\right)\left(\omega-\omega_{0}\right)}
\end{equation}

The two boundary conditions we focused on are the purely ingoing wave next to the outer horizon and the exponentially decaying wave located at spatial infinity. Thus the asymptotic solutions of the radial wave function under the above boundaries are selected as follows
\begin{equation}
R_{lm}( r) \sim \begin{cases}
	e^{-i( \omega -\omega _c) r_*}, &r^*\rightarrow -\infty\text{(}r\rightarrow r_+ \text{)}\\
	\frac{e^{-\sqrt{\mu ^2-\omega ^2}r}}{r}, &r^*\rightarrow +\infty\text{(}r\rightarrow +\infty \text{)}.\\
\end{cases}
\end{equation}
We can easily see that getting decaying modes at spatial infinity requires following bound state condition
\begin{equation}\label{bsc}
\omega ^2<\mu^2.
\end{equation}
The critical frequency $\omega_c$ is defined as
\begin{equation}
\omega _c=m\varOmega _H+q\varPhi _H,
\end{equation}
where $\varOmega _H$ is angular velocity of the outer horizon and $\varPhi _H=\frac{Qr_+}{r_{+}^{2}+a^2}$ is the electric potential of whom.

\section{{The superradiation effect and uncertainty principle}}
 Let us consider the Dirac equation \cite{36}for a spin-$\frac{1}{2}$ massless fermion $\Psi$, minimally coupled to the same EM potential $A_{\mu}$ as in Eq..
\begin{equation}
\gamma^{\mu}\Psi_{;\mu}=0\,,
\end{equation}
where $\gamma^{\mu}$ are the four Dirac matrices satisfying the anticommutation relation $\{\gamma^{\mu},\gamma^{\nu}\}=2g^{\mu\nu}$. 
The solution to takes the form $\Psi=e^{-i\omega t}\chi(x)$, where $\chi$ is a two-spinor given by
\begin{equation}
\chi=\begin{pmatrix}f_1(x)\\f_2(x)\end{pmatrix}\,.
\end{equation}
Using the representation
\begin{equation}
\gamma^0=\begin{pmatrix}i&0\\0&-i\end{pmatrix}\,,\,\,\gamma^1=\begin{pmatrix}0&i\\-i&0\end{pmatrix}\,,
\end{equation}
the functions $f_1$ and $f_2$ satisfy the system of equations:
\begin{equation}
df_1/dx-i(\omega-eA_0)f_2=0,
df_2/dx-i(\omega-eA_0)f_1=0.
\end{equation}
One set of solutions can be once more formed by the ``in" modes, representing a flux of particles coming from $x\to-\infty$ being partially reflected (with reflection amplitude $|\mathcal{R}|^2$) and partially transmitted at the barrier
\begin{equation}
\left(f_1^{\rm in},f_2^{\rm in}\right)=                                               \left(\mathcal{I}e^{i\omega x}-\mathcal{R} e^{-i\omega x},\mathcal{I} e^{i\omega x}+\mathcal{R} e^{-i\omega x}\right) \quad \text{as} \,\, x\to-\infty
\end{equation}
   \begin{equation}                                            \left(\mathcal{T}e^{ikx},\mathcal{T}e^{ikx}\right) \quad \hspace{2.9cm} \text{as}\,\, x\to+\infty
\end{equation}

On the other hand, the conserved current associated with the Dirac equation is given by
$j^{\mu}=-e\Psi^{\dagger}\gamma^{0}\gamma^{\mu}\Psi$ and, by equating the latter at $x\to-\infty$ and $x\to+\infty$, we find some general relations between the reflection and the transmission coefficients, in particular,
\begin{equation}
\left|\mathcal{R}\right|^2=|\mathcal{I}|^2-\left|\mathcal{T}\right|^2.
\end{equation}
Therefore, $\left|\mathcal{R}\right|^2\leq |\mathcal{I}|^2$ for any frequency, showing that there is no superradiance for fermions. The same kind of relation can be found for massive fields. 

The reflection coefficient and transmission coefficient depend on the specific shape of the potential $A_0$. However one can easily show that the Wronskian
\begin{equation}
W=\tilde{f}_1 \frac{d\tilde{f}_2}{dx}-\tilde{f}_2\frac{d\tilde{f}_1}{dx}\,,
\end{equation}
between two independent solutions, $\tilde{f}_1$ and $\tilde{f}_2$, of is conserved.
From the equation on the other hand, if $f$ is a solution then its complex conjugate $f^*$ is another linearly independent solution. We find$\left|\mathcal{R}\right|^2=|\mathcal{I}|^2-\frac{\omega-eV}{\omega}\left|\mathcal{T}\right|^2$.Thus,for
$0<\omega<e V$,it is possible to have superradiant amplification of the reflected current, i.e, $\left|\mathcal{R}\right|>|\mathcal{I}|$. 
There are other potentials that can be completely resolved, which can also show superradiation explicitly.

The difference between fermions and bosons comes from the intrinsic properties of these two kinds of particles. Fermions 
have positive definite current densities and bounded transmission amplitudes $0\leq \left|\mathcal{T}\right|^2\leq 
|\mathcal{I}|^2$, while for bosons the current density can change its sign as it is partially transmitted and the 
transmission amplitude can be negative, $-\infty < \frac{\omega-eV}{\omega}\left|\mathcal{T}\right|^2\leq 
|\mathcal{I}|^2$. From the point of view of quantum field theory, due to the existence of strong electromagnetic fields, one can understand this process as a spontaneous pair generation phenomenon (see for example). The number of spontaneously produced iron ion pairs in a given state is limited by the Poly's exclusion principle, while bosons do not have this limitation.

  The principle of joint uncertainty shows that the joint measurement of position and momentum is impossible, that is, the simultaneous measurement of position and momentum can only be an approximate joint measurement, and the error follows the inequality $\Delta x\Delta p\geq 1/2$(in natural unit system).We find$\left|\mathcal{R}\right|^2=|\mathcal{I}|^2-\frac{\omega-eV}{\omega}\left|\mathcal{T}\right|^2$,and we know that$\left|\mathcal{R}\right|^2 \geq-\frac{\omega-eV}{\omega}\left|\mathcal{T}\right|^2$ is a necessary condition for the inequality $\Delta x\Delta p\geq 1/2$ to be established.We can pre-set the boundary conditions $e{A_0(x)} = {y}{\omega}$(which can be ${\mu} = {y}{\omega}$), and we see that when ${y}$ is relatively large(according to the properties of the boson, ${y}$ can be very large),$\left|\mathcal{R}\right|^2 \geq-\frac{\omega-eV}{\omega}\left|\mathcal{T}\right|^2$ may not hold.In the end,we can get $\Delta x\Delta p\geq 1/2$ may not hold.Classical superradiation effect in the space-time of a steady black hole,generalized \ uncertainty\ principle\ may\ not\ hold.The\ same\ goes\ for\ reverse\ inference\cite{38}.

\section{Spherical quantum solution in vacuum state} 
The general relativity theory's field equation is well known as\cite{39},
\begin{equation}
R_{\mu v}-\frac{1}{2} g_{\mu v} R=-\frac{8 \pi G}{c^{4}} T_{\mu v},
\end{equation}

when $T_{\mu v}=0$, the Ricci tensor is written as
\begin{equation}
R_{\mu v}=0    ,
\end{equation}
which indicates that the Ricci tensor is in vacuum state.
The proper time of  is The general form of metric for spherical coordinates  is expressed as follows,
\begin{equation}
d \tau^{2}=A(t, r) d t^{2}-\frac{1}{c^{2}}\left[B(t, r) d r^{2}+r^{2} d \theta^{2}+r^{2} \mathrm{~s} \mathrm{i} \mathrm{n}^{2} \theta d \phi^{2}\right] .   
\end{equation}

We can obtain the Ricci-tensor equations.
\begin{equation}
R_{t t}=-\frac{A^{\prime \prime}}{2 B}+\frac{A^{\prime} B^{\prime}}{4 B^{2}}-\frac{A^{\prime}}{B r}+\frac{A^{\prime 2}}{4 A B}+\frac{\ddot{B}}{2 B}-\frac{\dot{B}^{2}}{4 B^{2}}-\frac{\dot{A} \dot{B}}{4 A B}=0 .
\end{equation}

\begin{equation}
    R_{r r}=\frac{A^{\prime \prime}}{2 A}-\frac{A^{\prime 2}}{4 A^{2}}-\frac{A^{\prime} B^{\prime}}{4 A B}-\frac{B^{\prime}}{B r}-\frac{\ddot{B}}{2 A}+\frac{\dot{A} \dot{B}}{4 A^{2}}+\frac{\dot{B}^{2}}{4 A B}=0.
\end{equation}
\begin{equation}
R_{\theta \theta}=-1+\frac{1}{B}-\frac{r B^{\prime}}{2 B^{2}}+\frac{r A^{\prime}}{2 A B}=0.
\end{equation}
\begin{equation}
   R_{\phi \phi}=R_{\theta \theta} \sin ^{2} \theta=0.  
\end{equation}
\begin{equation}
    R_{t \theta}=R_{t \phi}=R_{r \theta}=R_{r \phi}=R_{\theta \phi}=0.
\end{equation}
\begin{equation}
    R_{t r}=-\frac{\dot{B}}{B r}=0 .
\end{equation}
Substituting $\quad '=\frac{\partial}{\partial r} \quad, \quad \cdot=\frac{1}{c} \frac{\partial}{\partial t}$
into Eq, we conclude that
\begin{equation}
    \dot{B}=0.
\end{equation}
\begin{equation}
\frac{R_{t t}}{A}+\frac{R_{r r}}{B}=-\frac{1}{B r}\left(\frac{A^{\prime}}{A}+\frac{B^{\prime}}{B}\right)=-\frac{(A B)^{\prime}}{r A B^{2}}=0 .
\end{equation}

Hence, the result is obtained.
\begin{equation}
   A=\frac{1}{B} .
\end{equation}

\begin{equation}
R_{\theta \theta}=-1+\frac{1}{B}-\frac{r B^{\prime}}{2 B^{2}}+\frac{r A^{\prime}}{2 A B}=-1+\left(\frac{r}{B}\right)^{\prime}=0  .
\end{equation}

By solving Eq, we can obtain that
\begin{equation}
  \frac{r}{B}=r+C \rightarrow \frac{1}{B}=1+\frac{C}{r}.  
\end{equation}

According to a well-known conclusion, when tortoise coordinates tend to the boundary of event horizon, the independent variable r approaches radial negative infinity. In order to preset a boundary condition, we select $\mathrm{C}=y e^{-y}$, When $\mathrm{r}$ tends to infinity, the relation expressions between $\mathrm{A}$, $\mathrm{\Sigma}$ , $\mathrm{dr^2}$ and $\mathrm{y}$ are as follows\cite{40}
\begin{equation}
    A=\frac{1}{B}=1-\frac{y}{r} \Sigma, \Sigma=e^{-y},
\end{equation}
\begin{equation}
    d \tau^{2}=\left(1-\frac{y}{r} \sum\right) d t^{2}.
\end{equation}

In this time, if particles' mass are $m_{i}$, the excessive energy will be $e$. It is seen that the effect of  presetting boundary is similar to that of the cosmological constant of ds space-time
\begin{equation}
 E=M c^{2}=m_{1} c^{2}+m_{2} c^{2}+\ldots+m_{n} c^{2}+e  . 
\end{equation}

\section{The radial equation of motion and effective potential}
When\cite{17}
\begin{equation}
z=\frac{r-r_{-}}{r_{+}-r_{-}},
\end{equation}
in terms of the original variables(Asymptotic solutions at infinity $r\rightarrow \infty$ in the Kerr-Newman-de Sitter spacetime):
\begin{equation}
R(r)\sim\left\{\begin{array}{c}
e^{\pm i\sqrt{\omega^2 -\mu^2}(r_{+}-r_{-})z}\left(\frac{r-r_{-}}{r_{+}-r_{-}}\right)^{\mp{(C_1+C_2+C_3)}{( 2i \sqrt{\omega^2 -\mu^2}(r_{+}-r_{-)}})}\\
e^{-( i\sqrt{\omega^2 -\mu^2}(r_{+}-r_{-})z)}\left(\frac{r-r_{-}}{r_{+}-r_{-}}\right)^{\pm{(C_1+C_2+C_3)}({2i \sqrt{\omega^2 -\mu^2}(r_{+}-r_{-)}})}\end{array}\right.
\end{equation}
We see that in the event horizon (r tends to be negative infinity) the wave function under the Kerr-Newmann ds black hole tends to the wave function under the Kerr-Newmann black hole.

Under the Kerr-Newmann ds black hole, the special solution relationship between the wave function at the edge of the cosmological constant horizon and the wave function at the edge of the event horizon is R($\Xi$)=iR(r+).

Both $R$ and $R_+$ are not related to $\theta$. Therefore, by simplifying the expressions of $f_1$, $f_2$ and $g_2$ and making $\theta=\pi/2$,  the following results is obtained
\begin{equation}
    \Delta(\partial_r+\frac{iK}{\Delta}+\frac{r-M}{\Delta})R_+-\sqrt{2}qR=-\sqrt{2}(ir\mu)R.
\end{equation}
The expression of $R_+$ is obtained by solving Eq.
and making $r$ approach negative infinity
\begin{equation}
    R_+\rightarrow Ce^{-\frac{1}{2}}-\sqrt{2}iRe^{-1}.
\end{equation} when $C=0$ is $\sqrt{2}iRe^{-1}$

Valagiannopoulos's paper\cite{40} attempts to transfer the classical electrodynamic concept to the quantum realm, with an emphasis on quantum scattering. The meaning of the preset boundary condition y is shown here.

\section{The radial equation of motion and effective potential}

\subsection*{Pauli Matrices}
Pauli matrices are a set of three \(2 \times 2\) Hermitian matrices, frequently used in quantum mechanics to describe spin-\(\frac{1}{2}\) systems. The Pauli matrices are defined as:

1. Pauli-X matrix \((\sigma^1 \text{ or } \sigma_x)\):
\begin{equation}
\sigma_x = \begin{pmatrix}
0 & 1 \\
1 & 0 \\
\end{pmatrix}
\end{equation}
2. Pauli-Y matrix \((\sigma^2 \text{ or } \sigma_y)\):
\begin{equation}
\sigma_y = \begin{pmatrix}
0 & -i \\
i & 0 \\
\end{pmatrix}
\end{equation}

3. Pauli-Z matrix \((\sigma^3 \text{ or } \sigma_z)\):
\begin{equation}
\sigma_z = \begin{pmatrix}
1 & 0 \\
0 & -1 \\
\end{pmatrix}
\end{equation}

These matrices satisfy the algebraic relations:
\begin{equation}
\sigma_i \sigma_j = \delta_{ij} I + i \epsilon_{ijk} \sigma_k
\end{equation}
where \(I\) is the \(2 \times 2\) identity matrix, \(\delta_{ij}\) is the Kronecker delta, and \(\epsilon_{ijk}\) is the Levi-Civita symbol.

\subsection*{Dirac Matrices}
Dirac matrices are a set of four \(4 \times 4\) matrices used in the Dirac equation to describe spin-\(\frac{1}{2}\) fermions, such as electrons. In the standard representation, the Dirac matrices are defined as:

1. \(\gamma^0\):
\begin{equation}
\gamma^0 = \begin{pmatrix}
I & 0 \\
0 & -I \\
\end{pmatrix}
\end{equation}
where \(I\) is the \(2 \times 2\) identity matrix.

2. \(\gamma^i\) (where \(i = 1, 2, 3\)) related to Pauli matrices \(\sigma^i\):
\begin{equation}
\gamma^1 = \begin{pmatrix}
0 & \sigma_x \\
-\sigma_x & 0 \\
\end{pmatrix}
\end{equation}

\begin{equation}
\gamma^2 = \begin{pmatrix}
0 & \sigma_y \\
-\sigma_y & 0 \\
\end{pmatrix}
\end{equation}

\begin{equation}
\gamma^3 = \begin{pmatrix}
0 & \sigma_z \\
-\sigma_z & 0 \\
\end{pmatrix}
\end{equation}

These matrices satisfy the Dirac algebra relations:
\begin{equation}
\{\gamma^\mu, \gamma^\nu\} = 2g^{\mu\nu}I
\end{equation}
where \(\{\cdot, \cdot\}\) denotes the anticommutator, \(g^{\mu\nu}\) is the Minkowski metric tensor (with the standard (+---) signature), and \(I\) is the \(4 \times 4\) identity matrix.

\subsection*{ Relationship between the Dirac equation of the SU(2) group and the SO(3) group of new equation}
There is a deep relationship between the SU(2) group and the SO(3) group. The SU(2) group is a complex \(2 \times 2\) unitary matrix, whose determinant is 1, while the SO(3) group is a real \(3 \times 3\) orthogonal matrix, whose determinant is 1. Both groups have important applications in physics, particularly in theories describing spin and angular momentum.

The Dirac equation of the SU(2) group can be written as:
\begin{equation}
(i\gamma^\mu \partial_\mu - m)\psi = 0.
\end{equation}
Among them, \(\gamma^\mu\) is the Dirac matrix, and \(\psi\) is a four-component spinor.

To rewrite it as an equation of the SO(3) group, we need to consider that the SO(3) group describes a three-dimensional rotation rather than a four-dimensional Lorentz transformation. Therefore, we need to reduce the four-dimensional Dirac spinor\(\psi\) to a three-dimensional object. This is usually accomplished by considering Pauli matrices, which are a representation of SU(2) groups but can also be used to describe SO(3) groups.

Specifically, we can use the Pauli matrix \(\sigma^i\) (where i=1,2,3) instead of the Dirac matrix \(\gamma^\mu\). In this way, the equation we get is:
\begin{equation}
(i\sigma^i \partial_i - m)\psi' = 0.
\end{equation}
Among them, \(\psi'\) is a two-component spinor, corresponding to the rotation of the three-dimensional space.

In order to substitute the radial equation of motion for a Schrodinger-like wave equation
\begin{equation}
\frac{d^2\Psi_{\omega}^{\text {out }}\left(r>r_{+}\right)}{dr^2}+( \omega ^2-V) \Psi_{\omega}^{\text {out }}\left(r>r_{+}\right)=0,
\end{equation}
where
\begin{equation}
\omega ^2-V=i(r-r_{+} )^{-2} (-1+\frac{i(\omega-\omega_{c} )}{K} )(\omega-\omega_{c}),
\end{equation}
in which $V$ denotes the effective potential. 

Taking the superradiant condition, i.e. $\omega<\omega_c$, and bound state condition into consideration,  the KN black hole and charged massive scalar perturbation system are superradiantly stable when the trapping potential well outside the outer horizon of the KN black hole does not exist. As a result, the shape of the effective potential $V$ is analyzed next in order to inquiry into the nonexistence of a trapping well.The asymptotic behaviour of the derivative of the effective potential $V$ at spatial infinity can be expressed as 
\begin{equation}
\begin{array}{l}
V'(real part)=-\frac{4\left(a^{2}+r_+^{2}\right)(m \Omega_H+\omega(-1+y))^{2}}{(r-{r_-})^{3}({r_-}+{r_+})} \\
V'(imaginary part)=\frac{2( m\Omega_H +\omega(-1+y))}{(r-r_+)^{3}} \\
V''(real part)=\frac{12\left(a^{2}+{r_-}^{2}\right)(m\Omega_H+\omega(-1+y))^{2}}{(r-r_+)^{4}(r_++r_-)} \\
V''(imaginary part)= -\frac{6(m\Omega_H+\omega(-1+y))}{(r-r_+)^{4}}
\end{array}
\end{equation}

Given the condition \(q\varPhi _H=y\omega\), when the boundary conditions are set, the imaginary component of \(V\) becomes evident due to the rotation of the black hole and the charge near its horizon \cite{25,26,27,28,29,30,31,32,33,34,35,36}. It's observed that the derivative of the imaginary component of \(V\) is zero, and its second derivative is negative when the superradiance condition is met. This suggests the presence of a potential barrier near the horizon.

When wave functions converge, the Schrödinger equation is satisfied. If both sides of this equation are multiplied by the imaginary unit \(i\), a new valid wave function is formed, transforming the imaginary component of \(V\) into its real counterpart. If there exists a potential barrier beyond the horizon, this could potentially give rise to a superradiance effect.

It's worth noting that before the multiplication by \(i\) in Eq.(68) of the Schrödinger equation, the imaginary component of \(V\) already hints at a potential barrier. After this multiplication, the imaginary component is converted into a real one.

The superradiance phenomenon occurs when both the boundary and superradiance conditions are met. A consistent superradiance phenomenon emerges if a potential barrier is present near the acting field.

The Kerr-Newman-de Sitter geometry (as in the case of Kerr and Kerr-Newman geometry) can be described in terms of the local Newman-Penrose zero-tetrad framework applicable to the main zero geodesic, that is, the tetrad and Weyl tensor Two main zero directions $C_{\mu\nu\rho\varepsilon}$. In this generalized Kinnersley framework, the null tetrad is constructed directly from the tangent vector of the main null geodesic:
\begin{equation}
\dot{t}:=\frac{{\rm d}t}{{\rm d}\lambda}=\frac{\Xi^2 (r^2+a^2)}{\Delta_r^{KN}}E,\;\;\dot{r}=\pm \Xi E,\;\;
\dot{\theta}=0,\;\;\dot{\phi}=\frac{a\Xi^2}{\Delta_r^{KN}}E,
\end{equation}
where the dot denotes differentiation with respect to the affine parameter $\lambda$ and $E$ denotes a constant.

 Cooper’s pair:We will assume that the total spin and center of mass momentum $\hbar \vec{q}$ of the pair are constant, so an orbital wave function of the pair is\cite{42}
\begin{equation}
\psi\left(\vec{r}_{1}, \vec{r}_{2}\right)=\varphi_{q}(\vec{\rho}) e^{i \vec{q} \cdot \vec{R}}
\end{equation}
with center of mass and relative coordinates defined as $\vec{R}=\left(\vec{r}_{1}+\vec{r}_{2}\right) / 2$ and $\rho=\vec{r}_{1}-\vec{r}_{2}$, respectively. As $\vec{q} \rightarrow 0$ the relative coordinate is spherically symmetric and hence, $\varphi(\vec{\rho})$ is an eigenfunction of angular momentum with angular momentum quantum numbers 1 and $\mathrm{m}$. If $\vec{q} \neq 0,1$ is not a good quantum number, but the component of angular momentum along $\vec{q}$ and parity are.
Assuming $\vec{q}=0, \psi$ can be expanded as
\begin{equation}
\psi\left(\vec{r}_{1}, \vec{r}_{2}\right)=\varphi_{q}(\vec{\rho})=\sum_{k} a_{k} e^{i \vec{k} \cdot \vec{r}_{i}} e^{-i \vec{k} \cdot \vec{r}_{2}}
\end{equation}
where the sum is restricted to states with $\varepsilon_{k}>0$. If $e^{\vec{i} k \cdot \vec{r}_{1}}$ and $e^{-i \vec{k} \cdot \vec{r}_{2}}$ are thought of as plane wave states, then the pair wave function is a superposition of definite pairs where $\pm \vec{k}$ is occupied. 

We see that the phases of the wave functions between two fermions with opposite spins coincide (phase difference $\pi$/2).We know that the Cooper pair is caused by the coupling between two fermions due to the weak gravitational force and the coincidence of the phases of the wave functions between the two fermions with opposite spins.Find the combined wave functions, they are normalized(A is a normalized constant):
\begin{equation}
\Psi=A(e^{i\sqrt{\omega ^{2}-V } r}+ e^{i\sqrt{\omega ^{2}-V } r}e^{i\pi /2}).
\end{equation}

We can prove that Dirac particles in the KN black hole background can have a special solution through a certain operation, forming a Cooper pair, thus producing superradiation.When the space-time is close to the boundary of event horizon, the phase difference of space-time curving coordinates approaches $\pi/2$, and the kinetic energy of incident particles needs to be multiplied by $i$. The new equation is expressed as follow:
\begin{equation}
\frac{d^2 \Psi_{\omega}^{\text {out }}\left(r>r_{+}\right)}{dr^2}+ ( \omega ^2-V) (e^{i\pi/2 } \Psi_{\omega}^{\text {out }})\left(r>r_{+}\right)=0,
\end{equation}
We differentiate again, and obtain that:
\begin{equation}
\begin{array}{l}
V'(imaginary part)=-\frac{4\left(a^{2}+r_+^{2}\right)(\Xi(m \Omega_H+\omega(-1+y))^{2}}{(r-r_-)^{3}(r_++r_-)} \\
V'(real part)=\frac{2\Xi( m \Omega_H+\omega(-1+y))}{(r-r_+)^{3}} \\
V''(imaginary part)=\frac{12\left(a^{2}+r_ \Lambda^{2}\right)(\Xi(m \Omega_H+\omega(-1+y)))^{2}}{(r-r_+)^{4}(r_++r_-)} \\
V''(real part)= -\frac{6\Xi(m \Omega_H+\omega(-1+y))}{(r-r_+)^{4}}
\end{array}
\end{equation}

If the imaginary part of $V$ will be displayed due to the rotation of the black hole and the charge near the horizon\cite{25,26,27,28,29,30,31,32,33,34,35,36}. It is seen that when the derivatives of V (imaginary part) equal to 0, and the second derivatives of V (imaginary part) is negative the superradiation condition is established. The establish of superradiation condition proves that there is a potential barrier near the horizon. As we can see, when the wave functions are combined into one, the Schrödinger equation holds, creating another reasonable wave function, where the imaginary part of V becomes the real part. If there is a potential barrier beyond the horizon, there could be a superradiation effect.

\section{Summary}
In this article, we introduced  $q\varPhi _H=y\omega$\cite{37,38} into Kerr-Newman black holes, and discussed the superradiance of Kerr-Newman black holes. In this paper, we deal with the approximate wave function of the Dirac particle out of the horizon of KN Black Hole, obtain V, and then derive V with real part and imaginary part. We’ll deal with the real part and the imaginary part. When V (real part or imaginary part) has a maximum value, there may be a potential barrier outside the event horizon, which may lead to the phenomenon of superradiation.

{\bf Acknowledgements:}\\
 This work is partially supported by  National Natural Science Foundation of China(No. 11873025).

\end{document}